\documentclass[11pt]{article}
\usepackage{amsbsy}
\newcommand{\beq}{\begin{equation}}
\newcommand{\eeq}{\end{equation}}
\newcommand{\bdm}{\begin{displaymath}}
\newcommand{\edm}{\end{displaymath}}
\newcommand{\beqr}{\begin{eqnarray}}
\newcommand{\eeqr}{\end{eqnarray}}
\newcommand{\beqrn}{\begin{eqnarray*}}
\newcommand{\eeqrn}{\end{eqnarray*}}
\def\l{\lambda}

\def\R{{\bf R}}

\def\C{{\bf C}}
\def\Oc{{\bf O}}
\def\He{{\bf H}}
\def\bchi{\boldsymbol{\chi}}
\def\pder#1#2{\frac{\partial #1}{\partial #2}}%--- partial derivative
\def\ppder#1#2#3{\frac{\partial^2 #1}{\partial #2\partial #3}}

\def\ve{\varepsilon}
\def\a{\alpha}
\def\k{\kappa}
\def\nn{\nonumber}

\begin{document}

\title{Quantum trigonometric Calogero-Sutherland model, irreducible characters    and Clebsch-Gordan series for the exceptional algebra $E_7$ }

\author{J. Fern\'andez N\'u\~{n}ez$^{\;\dagger}$, W. Garc\'{\i}a Fuertes$^{\;\dagger}$,  A.M. 
Perelomov$^{\;\ddagger\;}$\footnote{On leave of absence from the Institute for Theoretical and Experimental Physics, 117259, Moscow, Russia. Current E-mail address: perelomo@mpim-bonn.mpg.de}\\     {\normalsize {\it $^\dagger$ Departamento de F\'{\i}sica, Facultad de Ciencias, Universidad de Oviedo, E-33007 Oviedo, Spain}} \\{\normalsize {\it $^\ddagger$ Departamento de F\'isica Te\'orica, Universidad de Zaragoza, E-50009 Zaragoza, Spain}}}

\date{}

\maketitle

\begin{abstract}
We re-express the quantum Calogero-Sutherland model for the Lie algebra $E_7$ and the particular value of the coupling constant $\kappa=1$ by using the fundamental irreducible characters of the algebra as dynamical variables. For that, we need to develop a systematic procedure to obtain all the Clebsch-Gordan series required to perform the change of variables. We describe how the resulting quantum Hamiltonian operator can be used to compute more characters and Clebsch-Gordan series for this exceptional algebra.

\end{abstract}

\section{Introduction}
Integrable systems are important because they can be considered as  0-th order perturbative aproximations to non-integrable systems. By {\it integrability} we mean here integrability in the sense of Liouville, that is, the existence of a complete set of mutually commuting integrals of motion. During the three last decades of the past century, a plethora of highly nontrivial (classical and quantum) mechanical integrable systems 
were discovered, see \cite{ca01, pe90} for comprehensive reviews. Among these, the Calogero-Sutherland  models form a distinguished class. 
The first analysis of a system of this kind was performed by Calogero 
\cite{ca71} who studied, from the quantum standpoint, the dynamics on the 
infinite line of a set of particles interacting pairwise by rational 
plus quadratic potentials, and found that the problem was exactly 
solvable. Soon afterwards, 
Sutherland \cite{su72} arrived to similar results for the quantum problem 
on the circle, this time with trigonometric interaction; and later 
 Moser \cite{mo75} proved, in terms of Lax pairs, that the classical counterparts of these models also
enjoyed integrability.

The identification of the 
general scope of these discoveries came with the work 
of Olshanetsky and Perelomov \cite{op76,op77,op78}, who realized that it 
is possible to associate models of this kind to all the root systems 
of the simple Lie algebras, and that all these models are integrable, 
both in the classical and  the quantum framework \cite{op81,op83}, for interactions of the type rational (or inverse-square), $q^{-2}$; rational+quadratic, $q^{-2}+\omega^2q^2$; trigonometric, $\sin^{-2}q$; hyperbolic, $\sinh^{-2}q$; and the most general, given by the Weierstrass elliptic function $\mathcal{P}(q)$. 
Nowadays, there is a widespread interest in this kind of integrable 
systems, and many mathematical and physical applications for them 
have been found, see for instance \cite{dv00}. In Physics, we mention, among others, the  remarkable connection  established \cite{cas96,cas03} between the different Calogero-Sutherland models and the properties of the equations describing the physics of disordered wires (the DMPK equation); the results are in good agreement with the experimental observations.

The study of the form and properties of the Schr\"{o}dinger eigenfunctions for the quantum version of these models constitutes by itself an interesting line of research. In fact, these eigenfunctions have very rich mathematical properties. In particular, for the trigonometric case, if we tune the coupling constants to some especial values, the wave functions correspond to the characters of the simple Lie algebras, while if we select a different tuning, we can make them to coincide with zonal spherical functions. Thus, the Calogero-Sutherland theories provide us with a new tool for computing these quantities. In this spirit, we will describe in the present paper how to use the trigonometric Calogero-Sutherland model to obtain both particular characters and Clebsch-Gordan series for the exceptional Lie algebra $E_7$. The main point of our approach is to express the  Hamiltonian in a suitable set of independent variables, indeed the fundamental characters of $E_7$. The use of such kind of variables has been quite useful to solve the Schr\"{o}dinger equation for the models associated to some  algebras \cite{op76}, \cite{pe98a}-\cite{ngp051}.

The organization of the paper is as follows. Section 2 is a reminder of the properties of $E_7$ relevant for the contents of the paper. Section 3 describes the main properties of the Calogero-Sutherland models associated to root systems and explains how to find the Hamiltonian in the variables mentioned above. Section 4 gives some account of the computation of the Clebsch-Gordan series of $E_7$ needed to pass to the new variables. In Section 5 we present the Hamiltonian in these variables and describe its use for computing new characters and to reduce tensor products of representations. Some conclusions are given in Section 6, and finally, the appendices show some explicit results for characters and Clebsch-Gordan series of $E_7$.

\section{Summary of results on the Lie algebra $E_7$}
In this Section, we review some standard facts about the root and weight systems of the Lie algebra $E_7$, with the aim of fixing the notation and help the reader to follow the rest of the paper. More extensive and sound treatments of these topics can be found in many excellent textbooks, see for instance \cite{ov90,otros}. 

The complex Lie algebra $E_7$ has dimension 133 and rank 7, as the name suggests. From the geometrical point of view, it admits (with some subtleties, see \cite{baez}) an interpretation which extends the standard-one for the classical algebras: in the same way that these correspond to the isometries of projective spaces over the first three normed division algebras ---$SO(n+1)\simeq {\rm Isom}(\R P^n)$, $SU(n+1)\simeq {\rm Isom}({\C}P^n)$, $Sp(n+1)\simeq {\rm Isom}({\He}P^n)$--- $F_4$, $E_6$, $E_7$ and $E_8$ are the Lie algebras of the projective planes over extensions of the octonions, giving rise to the so-called ``magic square": $F_4\simeq {\rm Isom}({\Oc}P^2)$, $E_6\simeq {\rm Isom}[({\C}\otimes{\Oc})P^2]$,  $E_7\simeq {\rm Isom}[({\He}\otimes{\Oc})P^2]$, $E_8\simeq {\rm Isom}[({\Oc}\otimes{\Oc})P^2]$. 
%In Physics, the most remarkable role played by $E_6$ is in the heterotic ten-%dimensional $E_8\times E_8$ superstring theory when the extra six dimensions are %compactified to a manifold of $SU(3)$ holonomy: in such a case, one of the $E_8$ %breaks to an $E_6$ which gives the Grand Unification group of four-dimensional %physics \cite{ramond}.  

The Dynkin diagram of $E_7$, see Figure 1, 

\begin{center}
\begin{picture}(80,65)(-20,-8)
\put(0,0){\circle{8}}
\put(4,0){\line(1,0){30}}
\put(38,0){\circle{8}}
\put(42,0){\line(1,0){30}}
\put(76,0){\circle{8}}
\put(0,4){\line(0,1){30}}
\put(0,38){\circle{8}}
\put(-4,0){\line(-1,0){30}}
\put(-38,0){\circle{8}}
\put(-42,0){\line(-1,0){30}}
\put(-76,0){\circle{8}}
\put(80,0){\line(1,0){30}}
\put(114,0){\circle{8}}
\put(-81,-15){$\a_1$}
\put(-43,-15){$\a_3$}
\put(-5,-15){$\a_4$}
\put(33,-15){$\a_5$}
\put(71,-15){$\a_6$}
\put(-4,46){$\a_2$}
\put(110,-15){$\a_7$}
\end{picture}
\end{center}
\bigskip
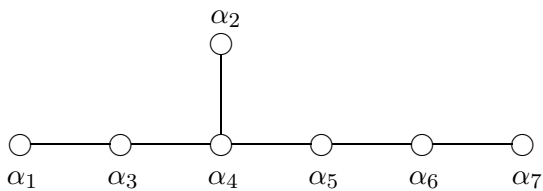
\centerline{ Figure 1. The Dynkin diagram for the Lie algebra $E_7$.}
\bigskip
encodes the euclidean relations $A_{ij}=(\a_i,\a_j)$ among the simple roots, which are
\beqr
\label{raices}
(\a_i,\a_i)&=&2,\hspace{1.5cm} i=1,\dots,7\nonumber\\
(\a_i,\a_{i+2})&=&-1,\hspace{1.2cm} i=1,2\\
(\a_i,\a_{i+1}) &=& -1,\hspace{1.2cm} i=3,\dots,6\nonumber\\
(\a_i,\a_j)&=&0, \hspace{1.2 cm}{\rm in\ all\ other\ cases}.\nonumber
\eeqr
Therefore, the Cartan matrix $A=(A_{ij})$ and its inverse $A^{-1}=(A^{-1}_{ij})$  read
\beq
\label{car}
A=\left(\begin{array}{ccccccc}
2&0&-1&0&0&0&0\\
0&2&0&-1&0&0&0\\
-1&0&2&-1&0&0&0\\
0&-1&-1&2&-1&0&0\\
0&0&0&-1&2&-1&0\\
0&0&0&0&-1&2&-1\\
0&0&0&0&0&-1&2
\end{array}\right)\,,\quad
 A^{-1}=\frac{1}{2}\left(\begin{array}{ccccccc}
4&4&6&8&6&4&2\\
4&7&8&12&9&6&3\\
6&8&12&16&12&8&4\\
8&12&16&24&18&12&6\\
6&9&12&18&15&10&5\\
4&6&8&12&10&8&4\\
2&3&4&6&5&4&3
\end{array}\right)\,.
\eeq

Throughout this paper we will use a realization of this  root system in terms of a  system of vectors $\{v_i\}_{i=1,\dots,8}$ of $\R^8$  (endowed with the standard Euclidean product $(\,,\,)$)  satisfying the relations $(v_i,v_j)=-\frac{1}{8}+\delta_{ij}$ \cite{ov90}. With reference to this system,  $E_7$  is the root system   in the hyperplane $V\subset\R^8$ of equation $\sum_iv_i=0$ given by ${\cal R}=\{v_i-v_j,v_i+v_j+v_k+v_l\,|\, i\ne j\ne k\ne l\}$, the positive ones being those in the subset  ${\cal R}^+=\{v_i-v_j,\ v_8-v_i,\ v_i+v_j+v_k+v_8\,|\,i<j<k<8\}$. There are 63 positive roots, which can be classified by heights as indicated in Table 1. The seven simple roots are
\begin{eqnarray}
\label{raep}
\a_1&=&v_1-v_2,\hspace{1.5cm} \a_2=v_4+v_5+v_6+v_7,\nonumber\\
\a_3&=&v_2-v_3,\hspace{1.5cm}\a_4=v_3-v_4\label{e61},\\
\a_5&=&v_4-v_5,\hspace{1.5cm}\a_6=v_5-v_6,\nonumber\\
\a_7&=&v_6-v_7,\nonumber
\end{eqnarray}
which clearly satisfy the relations (\ref{raices}).
\begin{table}[h]
\begin{center}
\begin{tabular}{|c|l|}
\hline
{\bf ht} & {\bf\hfil Positive roots} \\
\hline\hline
1 & $\a_1,\ \a_2,\ \a_3,\ \a_4,\ \a_5, \ \a_6,\ \a_7$ \\
\hline
2 & $\a_1+\a_3,\ \a_3+\a_4,\ \a_4+\a_5,\ \a_5+\a_6,\ \a_2+\a_4,\ \a_6+\a_7 $\\
\hline
3 & $\a_1+\a_3+\a_4,\ \a_3+\a_4+\a_5,\ \a_4+\a_5+\a_6,\ \a_2+\a_3+\a_4,\  \a_2+\a_4+\a_5,\ \a_5+\a_6+\a_7$\\
\hline
4 & $\a_1+\a_3+\a_4+\a_5,\ \a_3+\a_4+\a_5+\a_6,\ \a_1+\a_2+\a_3+\a_4,\  \a_2+\a_3+\a_4+\a_5,$\\ & $\ \a_2+\a_4+\a_5+\a_6,\ \a_4+\a_5+\a_6+\a_7$ \\
\hline
5 & $\a_1+\a_3+\a_4+\a_5+\a_6,\ \a_1+\a_2+\a_3+\a_4+\a_5,\ \a_2+\a_3+2\a_4+\a_5,$\\ &$\ \a_2+\a_3+\a_4+\a_5+\a_6,\ \a_2+\a_4+\a_5+\a_6+\a_7,\ \a_3+\a_4+\a_5+\a_6+\a_7$ \\
\hline
6 & $ \a_1+\a_2+\a_3+2\a_4+\a_5,\ \a_1+\a_2+\a_3+\a_4+\a_5+\a_6,\ \a_2+\a_3+2 \a_4+\a_5+\a_6,$\\ & $\ \a_1+\a_3+\a_4+\a_5+\a_6+\a_7,\ \a_2+\a_3+\a_4+\a_5+\a_6+\a_7$\\
\hline
7 & $\a_1+\a_2+2\a_3+2\a_4+\a_5,\ \a_2+\a_3+2\a_4+2 \a_5+\a_6,\  \a_1+\a_2+\a_3+2\a_4+\a_5+\a_6,$\\ &$\ \a_1+\a_2+\a_3+\a_4+\a_5+\a_6+\a_7,\ \a_2+\a_3+2\a_4+\a_5+\a_6+\a_7$ \\
\hline
8 & $\a_1+\a_2+2 \a_3+2 \a_4+\a_5+\a_6,\ \a_1+\a_2+\a_3+2 \a_4+2 \a_5+\a_6,$\\ &$\ \a_1+\a_2+\a_3+2\a_4+\a_5+\a_6+\a_7,\ \a_2+\a_3+2\a_4+2\a_5+\a_6+\a_7$ \\
\hline
9 & $\a_1+\a_2+2 \a_3+2 \a_4+2 \a_5+\a_6,\ \a_1+\a_2+\a_3+2\a_4+2\a_5+\a_6+\a_7,$\\ &$\ \a_1+\a_2+2\a_3+2\a_4+\a_5+\a_6+\a_7,\ \a_2+\a_3+2\a_4+2\a_5+2\a_6+\a_7 $ \\
\hline
10 & $\a_1+\a_2+\a_3+2\a_4+2\a_5+2\a_6+\a_7,\ \a_1+\a_2+2\a_3+2\a_4+2\a_5+\a_6+\a_7,$\\ &$\ \a_1+ \a_2+2 \a_3+3 \a_4+2 \a_5+\a_6$\\
\hline
11 & $\a_1+ 2\a_2+2 \a_3+3 \a_4+2 \a_5+\a_6,\ \a_1+\a_2+2\a_3+2\a_4+2\a_5+2\a_6+\a_7,$\\ &$\ \a_1+\a_2+2\a_3+3\a_4+2\a_5+\a_6+\a_7$\\
\hline
12 & $\a_1+\a_2+2\a_3+3\a_4+2\a_5+2\a_6+\a_7,\ \a_1+2\a_2+2\a_3+3\a_4+2\a_5+\a_6+\a_7$\\
\hline
13 & $\a_1+\a_2+2\a_3+3\a_4+3\a_5+2\a_6+\a_7,\ \a_1+2\a_2+2\a_3+3\a_4+2\a_5+2\a_6+\a_7$\\
\hline
14 & $\a_1+2\a_2+2\a_3+3\a_4+3\a_5+2\a_6+\a_7$\\
\hline
15 & $\a_1+2\a_2+2\a_3+4\a_4+3\a_5+2\a_6+\a_7$\\
\hline
16 & $\a_1+2\a_2+3\a_3+4\a_4+3\a_5+2\a_6+\a_7$\\
\hline
17 & $2\a_1+2\a_2+3\a_3+4\a_4+3\a_5+2\a_6+\a_7$\\
\hline
\end{tabular}
\caption[smallcaption]{Heights of positive roots of $E_7$.}
\end{center}
\end{table}

The hyperplane $V$ can be viewed as $\R^7$, and the basis made with the vectors $v_1,\dots,v_7$ is related to the canonical basis $\{e_k\}_{k=1,\dots,7}$ by $v_k=e_k-\frac{1}{7}(1+\frac{1}{\sqrt{8}})\sum_{j=1}^7e_j$; thus, the simple roots $\a_i$ are given by
\beqr
\a_1&=&e_1-e_2,\nn\\
\a_2&=&\frac{1}{7}\left(3-\frac{2}{\sqrt{2}}\right)\sum_{j=4}^7 e_j-\frac{4}{7}\left(1+\frac{1}{\sqrt{8}}\right)\sum_{j=1}^3 e_j,\\
\a_k&=&e_{k-1}-e_k,\ \ \ \ \ k=3,\dots,7.\nn
\eeqr

The fundamental weights $\l_i=\sum_{j=1}^7 A^{-1}_{ji}\a_j$ are
\beqrn
\l_1&=&v_1-v_8,\\
\l_2&=&-2v_8,\\
\l_3&=&v_1+v_2-2v_8,\\
\l_4&=&v_1+v_2+v_3+3v-3v_8,\\
\l_5&=&v_1+v_2+v_3+v_4-2v_8,\\
\l_6&=&v_1+v_2+v_3+v_4+v_5 -v_8,\\
\l_7&=&v_1+v_2+v_3+v_4+v_5+v_6 ,
\eeqrn
as it follows  from   (\ref{car}) and  (\ref{raep}).  As $E_7$ is simply-laced, the geometry of the weight system is summarized by the relations $(\l_i,\l_j)=A_{ij}^{-1}$. The Weyl vector is
\bdm
\rho =\frac{1}{2}\sum_{\a\in {\cal R}^+}\a=\sum_{i=1}^7 \l_i=\frac{1}{2}(34\a_1+49\a_2+66\a_3+96\a_4+75\a_5+52\a_6+27\a_7),
\edm
and  has  lenght  $|\rho|=\sqrt{798}/2$. The Weyl formula for dimensions applied to the irreducible representation associated to the integral dominant weight $\mu=\sum_{i=1}^7m_i\l_i$ gives
\bdm
\dim R_\mu=\prod_{\a\in {\cal R}^+} \frac{(\a, \mu+\rho)}{(\a,\rho)}=\frac{P}{2^6\cdot 3^6\cdot 4^6\cdot 5^6 \cdot 6^5 \cdot 7^5\cdot 8^4 \cdot 9^4\cdot 10^3\cdot 11^3\cdot12^2\cdot13^2\cdot14\cdot15\cdot16\cdot17},
\edm
where $P$ is a product extended to the set of positive roots in which the root $\a=\sum_{i=1}^7 a_i\,\a_i$ contributes with a factor ${\rm ht}(\a)+\sum_{i=1}^7 a_i\, m_i$, where ${\rm ht}(\a)$ is the height of $\a$.
In particular, for the basic representations $R_{\l_k}$, one finds:
\beqrn
&&\dim R_{\l_1}=133,\qquad\quad\dim R_{\l_2}=912,\\
&&\dim R_{\l_3}=8645,\qquad\ \, \dim R_{\l_4}=365750,\\
&&\dim R_{\l_5}=27664,\qquad\dim R_{\l_6}=1539,\\
&&\dim R_{\l_7}=56 .
\eeqrn
All the irreducible representations  are self-adjoint;  $R_{\l_1}$ is the adjoint representation  and $R_{\l_7}$, the fundamental one.

\section{The trigonometric Calogero-Sutherland model associated to a root system}
First of all, we review briefly the general theory of the quantum trigonometric Calogero-Sutherland model related to a root system $\cal R$ associated to a simple Lie algebra $L$ of rank $r$, and later  study explicity the $E_7$ case. For Calogero-Sutherland systems other than trigonometric see \cite{op83}; see also \cite{sa00}.

The  trigonometric Calogero-Sutherland model related to the root system $\cal R$ of rank $r$ is the quantum system in an Euclidean space $\R^r$ defined by the standard Hamiltonian operator 
\beq
\label{ham}
H=\frac{1}{2}\sum_{j=1}^rp_j ^2+\sum_{\a\in{\cal R}^+}\kappa_\a(\kappa_\a-1)\sin^{-2}(\a,q),
\eeq
where $q=({q_j})$ is  a cartesian coordinate system and $p_j=-{\rm i}\,\partial_{q_j}$; ${\cal R}^+$  is the set of the positive roots of $L$, and the coupling constants $\k_\a$ are such that $\kappa_\a=\kappa_\beta$ if $|\a|=|\beta|$. We will restrict ourselves to the case of simply-laced root systems (as the $E$-series is),  for which the Calogero-Sutherland model  depends only on one coupling constant $\k$. 

Although the Hamiltonian (\ref{ham}) is defined in all $\R^r$, the configuration space is confined by the singularities (infinite walls) $(\a,q)=0$. If the $q$-coordinates are assumed to take values in the $[0,\pi]$ interval, $H$ can be interpreted as describing the dynamics of a system of $r$ unit mass  particles   moving on the circle with interaction $V(q)=\k(\k-1)\sum_{\a}\sin^{-2}(\a,q)$, but notice that there is not translational invariance. The wave functions have to be $\pi$-periodic.

The main problem is to find the stationary states, i.e., to solve the Schr\"odinger eigenvalue problem $H\Psi=E\Psi$. The following important facts about this family of  quantum mechanical systems were well established in \cite{op76,op83}. 

(a)They are integrable, and moreover they are exactly solvable. The configuration space is confined to the Weyl alcove $\Lambda_W=\{q\in\R^r\,|\,0<(\a,q)<\pi\}$.

(b)The ground state energy and (non-normalized) wave function are
\begin{eqnarray*}
E_0(\k)&=&2 \rho^2\k^2\nn\\
\Psi_0^\k(q)&=&{\prod_{\a\in {\cal R}^+}\sin^\k(\a, q)},
\end{eqnarray*}
 while the excited states are indexed by the highest weights $\mu=\sum m_i\l_i\in P^+$ ($P^+$ is the cone of dominant weights) of the irreducible representations of $L$, that is, by the $r$-tuple of non-negative integers ${\bf m}=(m_1,\dots,m_r)$ (the quantum numbers), and the wave functions satisfy
\begin{eqnarray}
H\Psi^\k_{\bf m}&=&E_{\bf m}(\k)\Psi_{\bf m}^\k\nn\\
E_{\bf m}(\k)&=&2 (\mu+\k\rho,\mu+\k\rho).\label{105}
\end{eqnarray}

(c)It is natural to look for the solutions $\Psi_{\bf m}^\k$ in the form 
\beq
\Psi_{\bf m}^\k(q)=\Psi_0^\k(q)\Phi_{\bf m}^\k(q),
\eeq
and consequently we are led to the eigenvalue problem
\beq
\Delta^\k\Phi_{\bf m}^\k=\ve_{\bf m}(\k)\Phi_{\bf m}^\k\,,
\label{sch}
\eeq
where $\Delta ^\k$ is the linear differential operator
\beq
\Delta^\k=-\frac{1}{2}\sum_{j=1}^r\partial_{q_j}^{\,2}-\k\sum_{\a\in {\cal R}^+}  {\cot}(\a, q)(\a,\nabla_q)
\label{4b},
\eeq
and the eigenvalues $\ve_{\bf m}(\k)$ are the energies over the ground level, i.e.,
\beq
\label{energ}
\ve_{\bf m}(\k)=E_{\bf m}(\k)-E_0(\k)= 2(\mu, \mu+2\k\rho).
\eeq
Taking into account that $(\l_j,\l_k)=A_{jk}^{-1}$, it is possible to give a more explicit expression for the eigenvalues $\varepsilon_{\bf m}(\kappa)$:
\beq
\ve_{\bf m}(\k)=2\sum_{j,k=1}^r A_{jk}^{-1} m_j m_k+4\k\sum_{j,k=1}^r A_{jk}^{-1}m_j.
\label{eigenvalues}
\eeq
We will write $\ve_j(\k)$ for the fundamental weigth $\l_j$, i.e., for the quantum numbers   $(0,\dots,\buildrel(j)\over1,\dots,0)$

(d)In the case $\k=0$ the wave functions (\ref{sch}) are (proportional to) the monomial symmetric functions
\beq
M_\l(q)=\sum_{w\in W}e^{2i(w\cdot \l,q)},\ \l\in P^+\,,
\eeq
$W$ being the Weyl group of $L$. And the wave functions in the case $\k=1$ are (proportional to) the characters of the irreducible representations
\beq
\bchi_\l(q)=\frac{\sum_{w\in W}(\det w)e^{2i(w\cdot(\l+\rho),q)}}{\sum_{w\in W}(\det w)e^{2i(w\cdot \rho,q)}},\ \l\in P^+\,.
\eeq
Both $M_\l$ and $\bchi_\l$ are sums over the orbit of $\l$ under $W$, and consequently, $W$-invariant; as wave functions, they represent  superpositions of plane waves whose momenta are consistent with the required $\pi$-periodicity.

(d)Due to the Weyl symmetry of the Hamiltonian, the wave functions $\Phi_{\bf m}^\k(q)$ are   $W$-invariant, and the best way to solve the eigenvalue problem (\ref{sch}) is to use the set of independent $W$-invariant variables $z_k=\bchi_{\l_k}(q)$, in terms of which the wave functions $\Phi_{\bf m}^\k$ are polynomials.

Unfortunately, the expression of these characters $z_k$ in terms of the $q$-variables   is complicated and makes the direct change of variables $z=z(q)$ very cumbersome. We  are thus forced to follow a much more convenient, indirect route, which has proven to be useful for other root systems, \cite{ngp03,ngp051}. 

To this goal, the starting point is to write the operator $\Delta^\k$ in the $z$-variables:
\beq
\Delta^\kappa=\sum_{j, k} a_{jk}(z)\partial_{z_j}\partial_{z_k}+\sum_{j} \left[b_j^{0}(z)+\kappa\, b_j^{1}(z)\right]\partial_{z_j},
\label{deltaz}
\eeq
with $a_{jk}=a_{kj}$. Now,  if we take into account the  fact that, as pointed above, 
$b_j^{0}(z)+ b_j^{1}(z)=\Delta^1z_j=\ve_{j}(1)z_j $, the full expression for the coefficients $b_j(q)=b_j^{0}(z)+ b_j^{1}(z)$ appearing in $\Delta^1$ is completely determined by the Cartan matrix of the algebra;   explicitly 
\beq
\label{bes}
b_j(z)=2(A_{jj}^{-1}+2\sum_kA_{kj}^{-1})z_j,\ j=1,\dots,r.
\eeq

On the other hand, in order to find the coefficients $a_{jk}$ we will relay on the quadratic Clebsh-Gordan series 
\beq
\label{cbs}
R_{\l_j}\otimes R_{\l_k}=\sum_{\mu\in Q_{jk}}N_{\mu;jk}\,R_\mu,
\eeq
where $Q_{jk}\subset P^+$ is the  set of dominant weights  in the irreducible representation of highest weight $\l_j+\l_k$, and $N_{\l;jk}$ is the multiplicity of the irreducible representation $R_\l$ in that series; in particular,  $N_{\l_j+\l_k;jk}=1$. In these expressions we will write $\bf m$ or $(m_1,\dots,m_r)$ instead of $\mu=\sum_i m_i\l_i$. The Clebsh-Gordan series (\ref{cbs}) yield the formulas
\beq
\label{cgsz}
z_jz_k=\sum_{{\bf m}\in Q_{jk}}N_{{\bf m};jk}\,\bchi_{\bf m}(z)
\eeq
for the products of fundamental characters $z_jz_k$, and consequently we obtain the coefficients $a_{jk}$ by applying the operator $\Delta^1$ to the two members of (\ref{cgsz}):
\beq
2a_{jk}(z)=\sum_{{\bf m}\in Q_{jk}}N_{{\bf m};jk}\,\ve_{\bf m}(1)\,\bchi_{\bf m}(z)-b_j(z)z_k-b_k(z)z_j,\ j,k=1,\dots,r.\label{losajk}
\eeq

Therefore, to accomplish the task of fixing the form of the coefficents $a_{jk}$ we need  the list of all the quadratic Clebsh-Gordan series, the explicit expressions of the characters entering in them, and the coefficients $b_j$. Although there are some results for $E_7$ already available in the literature \cite{ov90, slan}, most of the required Clebsch-Gordan series and characters remain, to our knowledge, to be calculated.

The remaining step to achieve the complete expression of $\Delta^\k$ is to look for the coefficients $b^0_j(z)$. These can be found if we know enough monomial symmetric functions $M_\l$ in terms of the $z$-variables. Suppose that the relations $M_k=M_k(z)$ are known, where $M_k=M_{\l_k}, k=1,\dots,r$; then, from the eigenvalue equation $\Delta^0M_k=\ve_k(0)M_k$ we obtain the following linear system for the $b^0$'s:
\beq
\label{b0}
\sum_{i,j}a_{ij}(z)\ppder{M_k}{z_i}{z_j}+\sum_jb_j^0(z)\pder{M_k}{z_j}=2\l^2_kM_k(z).
\eeq
This system has a unique solution  $(b_j^0)$ because each of the sets of characters and  monomial symmetric functions constitutes a basis of $W$--invariant functions.

Recently \cite{ngp052} we have sound how to find the functions $M_k(z)$ in the $E_6$ case. In the present paper we will study only the case $\k=1$ and consequently we do not need to calculate the $b^0$--coefficients now.

\section{The quadratic Clebsh-Gordan series for $E_7$}

We have developed a systematic strategy, entirely based in a few elementary facts, to obtain all quantities needed for application of the formula (\ref{losajk}).  This strategy, which is essentially the same used in the previous paper \cite{ngp051} for the case of $E_6$, was described there in full detail, so we will confine ourselves here to mention some very general but important points. First of all, the series should be computed starting from those involving the most external dots of the Dynkin diagram, and going gradually towards the center of it. This is the order that allows the most efficient use of the orthogonality relations. Second, the orthogonality relations should be used not only to fix the multiplicity of some of the weights of lower height, but also to determine linear equations among the multiplicities of several weights of intermediate height. While for $E_6$ this is not of great importance, for the more complicated case of $E_7$ an extensive use of such linear constraints is required. This constraints, along with the bounds on multiplicities established in  \cite{wy}, make it posible to write a system of diophantine equations with unique solution for these multiplicities.
Finally, once all the series are found, the inversion of them to obtain the second-order characters appearing in (\ref{losajk}) requires the computation of many other characters of third, fourth and fifth order. The best way to perform these computations is as follows. Starting from the outer region of the Dynkin diagram, we build in each step the part of the $\Delta^1$ operator which only requires the characters that we already know. Then, we can use one of the the procedures described in Section 5 below to compute the characters needed to obtain the next coefficient through (\ref{losajk}), and so on. This is possible because (\ref{losajk}) gives each coefficient $a_{ij}(z)$ in terms of characters associated to weights whose height is lower or equal than $\lambda_i+\lambda_j$.

By means of these techniques, one finally arrives to the following list of Clebsch-Gordan series:

\begin{eqnarray*}
R_{\l_1}\otimes R_{\l_1} &=&R_{2\l_1} \oplus R_{\l_6} \oplus R_{\l_3} \oplus R_{\l_1} \oplus 1,\\
R_{\l_1} \otimes R_{\l_2} &=&R_{\l_1+\l_2} \oplus R_{\l_7} \oplus R_{\l_2} \oplus R_{\l_1+\l_7} \oplus R_{\l_5},\\
R_{\l_1} \otimes R_{\l_3} &=&R_{\l_1+\l_3} \oplus R_{\l_4} \oplus R_{\l_1} \oplus R_{\l_6} \oplus R_{\l_3} \oplus R_{2\l_1} \oplus R_{\l_1+\l_6} \oplus R_{\l_2+\l_7},\\
R_{\l_1} \otimes R_{\l_4} &=&R_{\l_1+\l_4} \oplus   R_{\l_2+\l_5} \oplus  R_{\l_3+\l_6} \oplus  R_{\l_1+\l_2+\l_7} \oplus R_{\l_5+\l_7} \oplus R_{2\l_2} \oplus R_{\l_1+ \l_3} \oplus  R_{\l_4} \oplus  R_{\l_1+\l_6}\\ 
&\oplus& R_{\l_2+\l_7} \oplus  R_{\l_3},\\
R_{\l_1} \otimes R_{\l_5} &=&R_{\l_1+\l_5} \oplus R_{\l_2+\l_6} \oplus R_{\l_2} \oplus R_{\l_5} \oplus R_{\l_1+\l_2} \oplus R_{\l_1+\l_7} \oplus R_{\l_3+\l_7} \oplus R_{\l_6+\l_7}\\
R_{\l_1} \otimes R_{\l_6} &=&R_{\l_1+\l_6} \oplus R_{\l_2+\l_7} \oplus R_{\l_3} \oplus R_{\l_6} \oplus R_{2\l_7}\oplus R_{\l_1}\\
R_{\l_1} \otimes R_{\l_7} &=&R_{\l_1+\l_7} \oplus R_{\l_2} \oplus R_{\l_7} \\
R_{\l_2} \otimes R_{\l_2} &=&R_{2\l_2}\oplus R_{\l_1} \oplus R_{2\l_7} \oplus R_{\l_6} \oplus R_{2\l_1} \oplus R_{\l_3}\oplus R_{\l_2+\l_7}\oplus R_{\l_1+\l_6}\oplus R_{\l_4}\oplus 1 \\
R_{\l_2} \otimes R_{\l_3} &=& R_{\l_2+\l_3} \oplus R_{\l_1+\l_5} \oplus R_{\l_2} \oplus R_{\l_5} \oplus R_{\l_7} \oplus R_{\l_2+\l_6} \oplus R_{\l_3+\l_7} \oplus R_{\l_6+\l_7} \oplus 2 R_{\l_1+\l_7} \\
&\oplus& 2 R_{\l_1+\l_2} \oplus R_{2\l_1+\l_7}  \\
R_{\l_2} \otimes R_{\l_4} &=& R_{\l_2+\l_4} \oplus  R_{\l_3+\l_5} \oplus R_{\l_1+\l_2+\l_6} \oplus  R_{2\l_2+\l_7} \oplus R_{\l_1+\l_3+\l_7} \oplus  R_{\l_5+\l_6} \oplus  R_{\l_4+\l_7} \oplus R_{\l_1+\l_6+\l_7}\\ 
&\oplus& R_{2\l_1+\l_2} \oplus    2 R_{\l_2+\l_3} \oplus  R_{\l_2+2\l_7} \oplus   2R_{\l_1+\l_5} \oplus   R_{2\l_1+\l_7} \oplus 2R_{\l_2+\l_6} \oplus   2R_{\l_3+\l_7} \\
&\oplus&  2R_{\l_1+\l_2} \oplus  R_{\l_6+\l_7} \oplus R_{\l_5} \oplus  R_{\l_1+\l_7} \oplus   R_{\l_2}\\
R_{\l_2} \otimes R_{\l_5} &=&R_{\l_2+\l_5} \oplus   R_{\l_3+\l_6} \oplus  R_{2\l_6} \oplus  R_{\l_1+\l_2+\l_7} \oplus R_{2\l_2} \oplus R_{\l_1+\l_3} \oplus R_{\l_5+ \l_7} \oplus  R_{\l_4} \oplus  R_{\l_1+2\l_7}\\ 
&\oplus& 2 R_{\l_1+\l_6} \oplus 2 R_{\l_2+\l_7} \oplus R_{2\l_7} \oplus R_{2\l_1} \oplus R_{\l_3}\oplus R_{\l_6} \oplus R_{\l_1} \\ 
R_{\l_2} \otimes R_{\l_6} &=&R_{\l_7} \oplus R_{\l_2} \oplus 2 R_{\l_1+\l_7} \oplus R_{\l_5} \oplus R_{\l_6+\l_7}\oplus R_{\l_1+\l_2}\oplus R_{\l_3+\l_7}\oplus  R_{\l_2+\l_6}\\
R_{\l_2}\otimes R_{\l_7} &=&R_{\l_2+\l_7}\oplus R_{\l_1} \oplus R_{\l_3} \oplus R_{\l_6} \\
R_{\l_3} \otimes R_{\l_3} &=&R_{2\l_3} \oplus R_{\l_1+\l_4} \oplus R_{\l_2+\l_5} \oplus R_{2\l_1+\l_6} \oplus  R_{\l_3+\l_6} \oplus 2 R_{\l_1+\l_2+\l_7} \oplus R_{2\l_6} \oplus R_{\l_5+\l_7} \oplus R_{3 \l_1}\\ 
&\oplus& R_{2\l_2}\oplus 2R_{\l_1+\l_3} \oplus R_{\l_1+2\l_7} 
\oplus 2 R_{\l_4} \oplus 3R_{\l_1+\l_6} \oplus 2R_{2\l_1} 
\oplus 2R_{\l_2+\l_7} \oplus R_{2\l_7} \oplus 2 R_{\l_3} \\
&\oplus& 2 R_{\l_6} \oplus R_{\l_1} \oplus 1 \\ 
R_{\l_3} \otimes R_{\l_4} &=&R_{\l_3+\l_4} \oplus  R_{\l_1+\l_2+\l_5} \oplus  R_{2\l_2+\l_6} \oplus  R_{\l_1+\l_3+\l_6} \oplus  R_{2\l_5} \oplus R_{\l_4+\l_6} \\&\oplus&   R_{2\l_1+\l_2+\l_7} \oplus R_{\l_1+2\l_6} \oplus  2 R_{\l_2+\l_3+\l_7} \oplus   2R_{\l_1+\l_5+\l_7} \oplus R_{2\l_1+\l_3} \oplus  R_{2\l_1+2\l_7} \\&\oplus& 2 R_{\l_1+2\l_2} \oplus R_{2\l_3} \oplus  3R_{\l_1+\l_4} \oplus  2R_{\l_2+\l_6+\l_7} \oplus 2R_{2\l_1+\l_6} \oplus  3R_{\l_2+\l_5} \oplus R_{\l_3+2\l_7} \oplus 4 R_{\l_3+\l_6}\\ &\oplus&  R_{\l_6+2\l_7} \oplus 5 R_{\l_1+\l_2+\l_7} \oplus   R_{2\l_6} \oplus  2R_{2\l_2} \oplus 3R_{\l_5+\l_7} \oplus  R_{3\l_1} \oplus  3R_{\l_1+\l_3} \oplus 2R_{\l_1+2\l_7} \oplus  3R_{\l_4} \\&\oplus&   4R_{\l_1+\l_6} \oplus R_{2\l_1} \oplus 3R_{\l_2+\l_7} \oplus R_{2\l_7} \oplus 2R_{\l_3} \oplus  R_{\l_6} \oplus R_{\l_1}         \\
R_{\l_3} \otimes R_{\l_5} &=& R_{\l_3+\l_5} \oplus   R_{\l_1+\l_2+\l_6} \oplus R_{2\l_2+\l_7} \oplus   R_{\l_1+\l_3+\l_7} \oplus  R_{\l_5+\l_6} \oplus  R_{\l_4+\l_7} \oplus 2R_{\l_1+\l_6+\l_7} \\ &\oplus& R_{2\l_1+\l_2}
\oplus R_{\l_2+2\l_7} \oplus 2 R_{\l_2+\l_3} \oplus  2R_{\l_1+\l_5} \oplus  R_{3\l_7} \oplus 2 R_{2\l_1+\l_7} \oplus 3 R_{\l_2+\l_6} \oplus  3 R_{\l_3+\l_7} \\&\oplus&    3 R_{\l_1+\l_2} \oplus 2R_{\l_6+\l_7} \oplus  2R_{\l_5} \oplus 3 R_{\l_1+\l_7} \oplus R_{\l_2} \oplus R_{\l_7} \\
R_{\l_3} \otimes R_{\l_6} &=& R_{\l_3+\l_6} \oplus   R_{\l_1+\l_2+\l_7} \oplus  R_{2\l_2} \oplus  R_{\l_1+\l_3} \oplus R_{\l_5+\l_7} \oplus R_{\l_4} \oplus R_{\l_1+2 \l_7} \oplus 2 R_{\l_1+\l_6} \oplus 2 R_{\l_2+\l_7}\\ 
&\oplus& R_{2\l_1} \oplus R_{2\l_7} \oplus 2 R_{\l_3} \oplus R_{\l_6} \oplus R_{\l_1} \\
R_{\l_3} \otimes R_{\l_7} &=&R_{\l_3+\l_7} \oplus R_{\l_1+\l_7} \oplus R_{\l_1+\l_2} \oplus R_{\l_2} \oplus R_{\l_5}\\
R_{\l_4} \otimes R_{\l_4}&=&R_{2\l_4}\oplus R_{\l_2+\l_3+\l_5}\oplus R_{2\l_3+\l_6}\oplus R_{\l_1+2\l_5}\oplus R_{\l_1+2\l_2+\l_6}\oplus R_{3\l_2+\l_7}\oplus R_{\l_1+\l_4+\l_6}\\
&\oplus& 2R_{\l_1+\l_2+\l_3+\l_7}\oplus 2R_{\l_2+\l_5+\l_6}\oplus 2R_{\l_2+\l_4+\l_7}\oplus R_{2\l_1+2\l_6}\oplus R_{2\l_1+2\l_2}\oplus R_{\l_3+2\l_6}\\
 &\oplus&     R_{2\l_1+\l_5+\l_7}\oplus 3R_{\l_3+\l_5+\l_7}\oplus R_{\l_1+2\l_3}\oplus 2R_{2\l_1+\l_4}\oplus 4R_{\l_1+\l_2+\l_6+\l_7}\oplus R_{3\l_6}\oplus 3R_{2\l_2+\l_3}\\
 &\oplus& R_{3\l_1+2\l_7}\oplus 2R_{\l_5+\l_6+\l_7}\oplus 2R_{2\l_2+2\l_7}\oplus 2R_{\l_1+\l_3+2\l_7}\oplus 3R_{\l_3+\l_4}\oplus 6R_{\l_1+\l_2+\l_5}\\
 &\oplus& 4R_{2\l_2+\l_6}\oplus 2R_{\l_4+2\l_7}\oplus 3R_{\l_1+\l_6+2\l_7} \oplus R_{3\l_1+\l_6}\oplus 2R_{2\l_5}\oplus 6R_{\l_1+\l_3+\l_6}\oplus 6R_{2\l_1+\l_2+\l_7}\\
 &\oplus& 7R_{\l_4+\l_6}\oplus R_{4\l_1}\oplus R_{\l_2+3\l_7}\oplus 4R_{\l_1+2\l_6}\oplus 9R_{\l_2+\l_3+\l_7}\oplus 9R_{\l_1+\l_5+\l_7}\oplus 3R_{2\l_1+\l_3}\\
&\oplus& 6R_{\l_1+2\l_2}\oplus 4R_{2\l_1+2\l_7}\oplus 4R_{2\l_3}\oplus 8R_{\l_2+\l_6+\l_7}\oplus 6R_{\l_3+2\l_7}\oplus 8R_{\l_1+\l_4}\oplus 8R_{\l_2+\l_5}\\
&\oplus& 8R_{2\l_1+\l_6}\oplus R_{4\l_7}\oplus 3R_{\l_6+2\l_7}\oplus 2R_{3\l_1}\oplus 9R_{\l_3+\l_6}\oplus 4R_{2\l_6} \oplus12R_{\l_1+\l_2+\l_7}\oplus 6R_{\l_1+\l_3}\\
 &\oplus& 3R_{2\l_2}\oplus 7R_{\l_5+\l_7}\oplus 7R_{\l_4}\oplus 6R_{\l_1+2\l_7}\oplus 7R_{\l_1+\l_6}\oplus 5R_{\l_2+\l_7}\oplus 2R_{2\l_7}\oplus 3R_{2\l_1}\oplus 3R_{\l_3} \\
 &\oplus&3R_{\l_6}\oplus R_{\l_1}\oplus 1\\
R_{\l_4} \otimes R_{\l_5} &=&R_{\l_4+\l_5} \oplus R_{\l_2+\l_3+\l_6} \oplus  R_{\l_1+\l_5+\l_6} \oplus  R_{2\l_3+\l_7} \oplus  R_{\l_1+2\l_2+\l_7} \oplus  R_{\l_1+\l_4+\l_7} \oplus  R_{\l_2+2\l_6} \\&\oplus&  2 R_{\l_2+\l_5+\l_7} \oplus 2 R_{\l_1+\l_2+\l_3} \oplus  R_{2\l_1+\l_6+\l_7} \oplus R_{3\l_2} \oplus 2R_{\l_3+\l_6+\l_7} \oplus 2R_{\l_2+\l_4} \\
&\oplus& R_{2\l_1+\l_5} \oplus  3R_{\l_3+\l_5} \oplus  2R_{\l_1+\l_2+2\l_7} \oplus   5R_{\l_1+\l_2+\l_6} \oplus R_{2\l_6+\l_7} \oplus  R_{\l_5+2\l_7} \oplus R_{3\l_1+\l_7} \\&\oplus&  4R_{\l_1+\l_3+\l_7} \oplus  3R_{\l_5+\l_6} \oplus  3R_{2\l_2+\l_7} \oplus  5R_{\l_4+\l_7} \oplus  3R_{2\l_1+\l_2} \oplus 5R_{\l_2+\l_3} \oplus  R_{\l_1+3\l_7}\\ &\oplus& 5R_{\l_1+\l_6+\l_7} \oplus  5R_{\l_1+\l_5} \oplus 3R_{\l_2+2\l_7} \oplus  5R_{\l_2+\l_6} \oplus  4R_{2\l_1+\l_7} \oplus 5R_{\l_3+\l_7} \oplus  4R_{\l_1+\l_2} \oplus  R_{3\l_7} \\&\oplus& 3R_{\l_6+\l_7}\oplus 3R_{\l_5}\oplus 3R_{\l_1+\l_7} \oplus R_{\l_2} \oplus  R_{\l_7}               \\
R_{\l_4} \otimes R_{\l_6} &=& R_{\l_4+\l_6} \oplus  R_{\l_2+\l_3+\l_7} \oplus   R_{2\l_3} \oplus   R_{\l_1+\l_5+\l_7} \oplus   R_{\l_2+\l_6+\l_7} \oplus  R_{\l_1+2\l_2} \oplus  R_{\l_1+\l_4} \oplus  R_{\l_3+2\l_7}\\ 
&\oplus& 2R_{\l_2+\l_5} \oplus  R_{2\l_1+\l_6} \oplus  2R_{\l_3+\l_6}\oplus         3R_{\l_1+\l_2+\l_7} \oplus  R_{2\l_6} \oplus 2 R_{\l_5+\l_7} \oplus   R_{2\l_2} \oplus R_{\l_1+2\l_7}\\& \oplus&   2R_{\l_1+\l_3} 
\oplus 3R_{\l_4} \oplus 2 R_{\l_1+\l_6} \oplus   2 R_{\l_2+\l_7} \oplus  R_{2\l_1} \oplus   R_{\l_3} \oplus  R_{\l_6} \\
R_{\l_4} \otimes R_{\l_7} &=& R_{\l_4+\l_7} \oplus   R_{\l_2+\l_3}\oplus R_{\l_1+\l_5} \oplus R_{\l_2+\l_6} \oplus R_{\l_3+\l_7} \oplus R_{\l_1+\l_2} \oplus R_{\l_5} \\ 
R_{\l_5} \otimes R_{\l_5} &=&R_{2\l_5} \oplus  R_{\l_4+\l_6} \oplus  R_{\l_2+\l_3+\l_7} \oplus R_{\l_1+2\l_6} \oplus  R_{2\l_3} \oplus   R_{\l_1+\l_5+\l_7} \oplus 2R_{\l_2+\l_6+\l_7} \\
&\oplus&   R_{\l_1+2\l_2} \oplus R_{2\l_1+2\l_7} \oplus   R_{\l_1+\l_4} \oplus  R_{\l_3+2\l_7} \oplus 2R_{\l_2+\l_5} \oplus  R_{2\l_1+\l_6} \oplus  3R_{\l_3+\l_6} \oplus 4R_{\l_1+\l_2+\l_7} \\
&\oplus&  R_{\l_6+2\l_7} \oplus R_{3\l_1} \oplus      2R_{2\l_6} \oplus 3R_{\l_5+\l_7} \oplus  2R_{2\l_2} \oplus 3R_{\l_1+2\l_7} \oplus  2R_{\l_1+\l_3}\\ 
&\oplus& 3R_{\l_4} \oplus 4R_{\l_1+\l_6} \oplus   3R_{\l_2+\l_7} \oplus 2 R_{2\l_7} \oplus   2R_{2\l_1} \oplus 2R_{\l_3} \oplus  2R_{\l_6} \oplus R_{\l_1} \oplus   1 \\
R_{\l_5} \otimes R_{\l_6} &=&R_{\l_5+\l_6} \oplus   R_{\l_4+\l_7}  \oplus R_{\l_1+\l_6+\l_7} \oplus R_{\l_2+ \l_3} \oplus  R_{\l_2+2 \l_7} \oplus  R_{\l_1+\l_5}\\ &\oplus& 2 R_{\l_2+\l_6} \oplus  R_{2\l_1+\l_7} \oplus 2 R_{\l_3+\l_7} \oplus 2 R_{\l_1+\l_2} \oplus 2R_{\l_6+\l_7}\oplus 2R_{\l_5} \oplus 2R_{\l_1+\l_7}\oplus R_{\l_2} \oplus R_{\l_7}  \\ 
R_{\l_5} \otimes R_{\l_7} &=&R_{\l_5+\l_7} \oplus R_{\l_4} \oplus R_{\l_6} \oplus R_{\l_3} \oplus R_{\l_1+\l_6} \oplus R_{\l_2+\l_7} \\
R_{\l_6} \otimes R_{\l_6} &=&R_{2\l_6} \oplus R_{\l_5+\l_7} \oplus  R_{\l_1} \oplus R_{\l_3} \oplus R_{\l_4} \oplus 2 R_{\l_6} \oplus R_{2\l_1} \oplus R_{2\l_7}\oplus R_{\l_1+\l_6}  \oplus 2 R_{\l_2+\l_7}\\ 
&\oplus& R_{\l_1+2\l_7} \oplus 1\\
R_{\l_6} \otimes R_{\l_7} &=&R_{\l_6+\l_7} \oplus R_{\l_5} \oplus R_{\l_1+\l_7} \oplus R_{\l_2} \oplus R_{\l_7}\\
R_{\l_7} \otimes R_{\l_7} &=&R_{2\l_7} \oplus R_{\l_1} \oplus R_{\l_6}  \oplus 1\\
\end{eqnarray*}

We present also a list of second order characters:
\begin{eqnarray*}
\bchi_{2000000} &=& z_1^2
-z_3
-z_6
-z_1
-1
\\
\bchi_{1100000} &=& z_1 z_2
-z_5
-z_1 z_7
\\
\bchi_{1010000} &=& z_1 z_3
-z_4
-z_1 z_6
-z_1^2
+z_7^2
+z_3
\\
\bchi_{1001000} &=& z_1 z_4
-z_2 z_5
+z_6^2
-z_5 z_7
+z_1 z_6
-z_2 z_7
-z_7^2
+z_6
+z_1
\\
\bchi_{1000100} &=& z_1 z_5
-z_2 z_6
+z_1 z_7
-z_2
\\
\bchi_{1000010} &=& z_1 z_6
-z_2 z_7
-z_7^2
+z_6
+z_1
+1
\\
\bchi_{1000001} &=& z_1 z_7
-z_2
-z_7
\\
\bchi_{0200000} &=& z_2^2
-z_4
-z_1 z_6
-z_1^2
+z_3
+z_6
+z_1
\\
\bchi_{0110000} &=& z_2 z_3
-z_1 z_5
-z_1^2 z_7
+z_3 z_7
+z_6 z_7
+z_1 z_7
\\
\bchi_{0101000} &=& z_2 z_4
-z_3 z_5
+z_1 z_6 z_7
-z_2 z_7^2
-z_1 z_5
+z_2 z_6
-z_6 z_7
+z_5
+z_2
\\
\bchi_{0100100} &=& z_2 z_5
-z_3 z_6
-z_6^2
+z_5 z_7
+z_1 z_7^2
-z_1 z_6
-z_6
-z_1
\\
\bchi_{0100010} &=& z_2 z_6
-z_3 z_7
-z_6 z_7
+z_5
+z_2
\\
\bchi_{0100001} &=& z_2 z_7
-z_3
-z_6
-z_1
\\
\bchi_{0020000} &=& z_3^2
-z_1 z_4
-z_1^2 z_6
+z_3 z_6
+z_5 z_7
-z_1^3
+2 z_1 z_3
+z_1 z_7^2
-z_4
+z_3
+z_1
\\
\bchi_{0011000} &=& z_3 z_4
-z_1 z_2 z_5
+z_4 z_6
+z_1 z_6^2
+z_1^2 z_6
-z_3 z_6
-z_6 z_7^2
-z_1 z_2 z_7
+z_2^2
+z_5 z_7
-z_4
-z_1 z_6
+z_7^2\\
&+&z_3
-1
\\
\bchi_{0010100} &=& z_3 z_5
-z_1 z_2 z_6
+z_4 z_7
+z_1 z_6 z_7
+z_2 z_7^2
-z_7^3
+z_1^2 z_7
-z_2 z_6
-z_3 z_7
-z_1 z_2
+z_6 z_7
-z_5
-z_1 z_7\\
&-&z_2
+z_7
\\
\bchi_{0010010} &=& z_3 z_6
-z_1 z_2 z_7
+z_4
+z_1 z_6
+z_2 z_7
+z_1^2
-z_3
-z_1
\\
\bchi_{0010001} &=& z_3 z_7
-z_1 z_2
+z_7
\\
\bchi_{0002000} &=& z_4^2
-z_2 z_3 z_5
+z_1 z_4 z_6
+z_3 z_6^2
+z_1^2 z_5 z_7
-2 z_3 z_5 z_7
-2 z_4 z_7^2
-z_1 z_6 z_7^2
-z_5^2
+2 z_4 z_6
+z_1 z_6^2
\\&-&z_1 z_5 z_7
+z_2 z_5
+z_7^4
-2 z_6 z_7^2
+z_6^2
+2 z_4
+z_1 z_6
-2 z_7^2
+2 z_6
+1
\\
\bchi_{0001100} &=& z_4 z_5
-z_2 z_3 z_6
+z_1 z_4 z_7
+z_1^2 z_6 z_7
-z_3 z_5
-z_5 z_6
-z_1 z_7^3
-z_4 z_7
-z_1 z_5
+z_7^3
-z_6 z_7
+z_1 z_7
-z_7
\\
\bchi_{0001010} &=& z_4 z_6
-z_2 z_3 z_7
+z_1^2 z_7^2
+z_1 z_4
-z_3 z_7^2
+z_3 z_6
-z_5 z_7
-2 z_1 z_7^2
+z_4
+z_1 z_6
-z_7^2
+z_6
+z_1
+1
\\
\bchi_{0001001} &=& z_4 z_7
-z_2 z_3
+z_1^2 z_7
-z_3 z_7
-z_5
-z_1 z_7
-z_7
\\
\bchi_{0000200} &=& z_5^2
-z_4 z_6
-z_1 z_6^2
+z_1 z_5 z_7
+z_3 z_7^2
-z_3 z_6
+z_6 z_7^2
-z_6^2
-z_4
-z_1 z_6
+z_7^2
-z_3
-2 z_6
-1
\\
\bchi_{0000110} &=& z_5 z_6
-z_4 z_7
-z_1 z_6 z_7
+z_1 z_5
+z_7^3
+z_3 z_7
-z_6 z_7
+z_5
-z_7
\\
\bchi_{0000101} &=& z_5 z_7
-z_4
-z_1 z_6
+z_7^2
-z_6
-1
\\
\bchi_{0000020} &=& z_6^2
-z_5 z_7
-z_1 z_7^2
+z_1 z_6
+z_3
+z_6
+z_1
\\
\bchi_{0000011} &=& z_6 z_7
-z_5
-z_1 z_7
\\
\bchi_{0000002} &=& z_7^2
-z_6
-z_1
-1
\end{eqnarray*}

\section{The Calogero-Sutherland Hamiltonian $\Delta^1$ in $E_7$. Some applications}

The  coefficients $b_j(z)$ in the expression of $\Delta^1$ are easily obtained from (\ref{bes}) and (\ref{car}):
\beqrn
&&b_1(z)=72z_1;\quad b_2(z)=105z_2;\quad b_3(z)=144z_1;\quad b_4(z)=216z_4;\\
 &&b_5(z)=165z_5;\quad b_6(z)=112z_5;\quad b_7(z)=57z_7\,.
\eeqrn

After having computed in Section 4 the necessary series and characters, we can now follow the lines indicated in Section 3 to obtain the Hamiltonian operator in the limit $\kappa=1$. The result for the coefficients $a_{jk}(z)$ in (\ref{deltaz}) for $\k=1$ is

\beqrn
a_{1  1} (z) &=& 4(-19 - 10\, z_1 + z_1^2 - z_3 - 5\,z_6)\\ 
a_{1  2} ( z) &=& 2(-7\, z_2 + 2\, z_1\, z_2 - 5\, z_5 - 19\, z_7 - 13\, z_1\, z_7)\\ 
a_{1  3} (z) &=& 2(10 - 14\, z_1 - 19\, z_1^2 + 13\, z_3 +3\, z_1\, z_3 - 3\, z_4 - 4\, z_6 - 9\, z_1\, z_6 - 6\, z_2\, z_7 +9\, z_7^2)\\ 
a_{1  4} ( z) &=& 2(-10 - 2\, z_1 + 18\, z_1^2 - 7\, z_2^2 - 24\, z_3 - 6\, z_1\, z_3 + 14\, z_4 + 4\, z_1\, z_4 - 4\, z_2\, z_5 + 8\, z_6 + 12\, z_1\, z_6 \\
      &-& 4\, z_3\, z_6 + 4\, z_6^2 - 14\, z_2\, z_7 - 5\, z_1\, z_2\, z_7 - 9\, z_5\, z_7 - 9\, z_7^2 + 9\, z_1\, z_7^2)\\ 
a_{1  5} ( z) &=& 2(-12\, z_2 - 6\, z_1\, z_2 + 8\, z_5 + 3\, z_1\, z_5 - 5\, z_2\, z_6 + 19\, z_7 + 5\, z_1\, z_7 - 5\, z_3\, z_7 - 13\, z_6\, z_7)\\ 
a_{1  6} (z) &=& 4(9 - 3\, z_1 - 3\, z_3 + 2\, z_6 + \, z_1\, z_6 - 3\, z_2\, z_7 - 9\, z_7^2)\\ 
a_{1  7} ( z) &=& 2(-7\, z_2 - 19\, z_7 + \, z_1\, z_7)\\ 
a_{2  2} (z) &=& -40 + 24\, z_1 - 36\, z_1^2 + 7\, z_2^2 + 24\, z_3 - 4\, z_4 + 44\, z_6 - 16\, z_1\, z_6 - 12\, z_2\, z_7 - 36\, z_7^2\\ 
a_{2  3} ( z) &=& 2(9\, z_2 - 14\, z_1\, z_2 + 4\, z_2\, z_3 + 16\, z_5 - 4\, z_1\, z_5 - 5\, z_2\, z_6 + 4\, z_1\, z_7 - 12\, z_1^2\, z_7 + 7\, z_3\, z_7 - \, z_6\, z_7)\\ 
a_{2  4} ( z) &=& 2(-7\, z_2 - 4\, z_1\, z_2 - 6\, z_1^2\, z_2 - \, z_2\, z_3 + 6\, z_2\, z_4 + 20\, z_5 - 6\, z_1\, z_5 - 3\, z_3\, z_5 + 14\, z_2\, z_6 - 4\, z_1\, z_2\, z_6 \\
	&-& 5\, z_5\, z_6 - 10\, z_1\, z_7 +22\, z_1^2\, z_7 - 6\, z_2^2\, z_7 - 12\, z_3\, z_7 - 5\, z_1\, z_3\, z_7 + 12\, z_4\, z_7\\
      &-& 10\, z_6\, z_7 + 13\, z_1\, z_6\, z_7 - 9\, z_2\, z_7^2)\\ 
a_{2  5} ( z) &=& 40 - 24\, z_1+ 12\, z_1^2 - 14\, z_2^2 + 12\, z_3 - 12\, z_1\, z_3 + 28\, z_4 + 9\, z_2\, z_5 + 16\, z_6 - 12\, z_1\, z_6 \\
	&-& 8\, z_3\, z_6 - 24\, z_6^2+ 12\, z_2\, z_7 - 10\, z_1\, z_2\, z_7 + 14\, z_5\, z_7 - 2\, z_7^2 - 2\, z_1\, z_7^2\\ 
a_{2  6} ( z) &=& 2(17\, z_2 - 6\, z_1\, z_2 + 8\, z_5 + 3\, z_2\, z_6 - 24\, z_1\, z_7 - 5\, z_3\, z_7 - 13\, z_6\, z_7)\\ 
a_{2  7} (z) &=& -48\, z_1 - 12\, z_3 - 28\, z_6 + 3\, z_2\, z_7\\ 
a_{3  3} ( z) &=& 4(-20 + 16\, z_1 - 5\, z_1^2 - 9\, z_1^3 + 8\, z_3 + 12\, z_1\, z_3 + 3\, z_3^2 -7\, z_4 - \, z_1\, z_4 - 2\, z_2\, z_5 - 9\, z_6 \\
      &+& -3\, z_1\, z_6 - 4\, z_1^2\, z_6+ 2\, z_3\, z_6 - 3\, z_6^2 + 6\, z_2\, z_7 - 6\, z_1\, z_2\, z_7 + 8\, z_5\, z_7 + \, z_7^2 + 7\, z_1\, z_7^2)\\ 
a_{3  4} ( z) &=& 2(-20 - 12\, z_1 - 6\, z_1^2 + 6\, z_1^3 + 9\, z_2^2 - 7\, z_1\, z_2^2 + 20\, z_3 + 6\, z_1\, z_3 - 6\, z_1^2\, z_3 + 6\, z_3^2 - 26\, z_4 + 6\, z_1\, z_4\\ 
      &+& 8\, z_3\, z_4 + 2\, z_2\, z_5 - 3\, z_1\, z_2\, z_5 - 5\, z_5^2 - 2\, z_6 - 32\, z_1\, z_6 + 20\, z_1^2\, z_6 - 5\, z_2^2\, z_6 - 24\, z_3\, z_6 - 4\, z_1\, z_3\, z_6\\ 
      &+& 10\, z_4\, z_6 + 4\, z_6^2 + 4\, z_1\, z_6^2 - 3\, z_2\, z_7 - 2\, z_1\, z_2\, z_7 - 5\, z_1^2\, z_2\, z_7 - \, z_2\, z_3\, z_7 + 15\, z_5\, z_7 + 4\, z_2\, z_6\, z_7\\ 
      &+& 20\, z_7^2 - 5\, z_1\, z_7^2 + 5\, z_1^2\, z_7^2 + 9\, z_3\, z_7^2 - 9\, z_6\, z_7^2)\\
a_{3  5} ( z) &=& 2(-\, z_2 + \, z_1\, z_2 - 6\, z_1^2\, z_2 - \, z_2\, z_3 - 22\, z_5 + 15\, z_1\, z_5 + 6\, z_3\, z_5 - 15\, z_2\, z_6 - 4\, z_1\, z_2\, z_6 - 5\, z_5\, z_6 \\
      &+& 18\, z_7- 5\, z_1\, z_7 + 11\, z_1^2\, z_7 - 6\, z_2^2\, z_7 - 6\, z_3\, z_7 - 5\, z_1\, z_3\, z_7 + 12\, z_4\, z_7 + 45\, z_6\, z_7 - 8\, z_1\, z_6\, z_7 \\
      &+& 12\, z_2\, z_7^2 - 18\, z_7^3)\\ 
a_{3  6} ( z) &=& 2(20 + 6\, z_1^2 - 7\, z_2^2 - 18\, z_3 - 6\, z_1\, z_3 + 14\, z_4 + 20\, z_6 + 6\, z_1\, z_6 + 4\, z_3\, z_6 + 6\, z_2\, z_7 - 5\, z_1\, z_2\, z_7\\ 
      &-& 5\, z_5\, z_7 - \, z_7^2 - 13\, z_1\, z_7^2) \\ 
a_{3  7} ( z) &=& 2(-7\, z_2 - 6\, z_1\, z_2 - 5\, z_5 + 19\, z_7 - 13\, z_1\, z_7 + 2\, z_3\, z_7)\\ 
a_{44}(z)&=&4(-10 - 16\, z_1 + 8\, z_1^2 - 6\, z_1^4 - 4\, z_2^2 - 16\, z_3+ 
    24\, z_1^2 z_3 - 4\, z_2^2 z_3 - 12\, z_3^2 + 8\, z_4 
    -16\, z_1 z_4 - 4\, z_1^2 z_4 \\
    &+& 8\, z_3 z_4 + 6\, z_4^2 + 
    11\, z_2 z_5 + 2\, z_1 z_2 z_5 -  z_2 z_3 z_5 - 12\, z_5^2 - 
    2\, z_1 z_5^2 + 4\, z_6 - 4\, z_1 z_6 - 14\, z_1^2 z_6 + 
    4\, z_1^3 z_6\\
    & +& 3\, z_2^2 z_6 - 2\, z_1 z_2^2 z_6 -    4\, z_3 z_6 - 8\, z_1 z_3 z_6 - 2\, z_3^2 z_6 + 6\, z_4 z_6 + 
    4\, z_1 z_4 z_6 - 3\, z_2 z_5 z_6 - 2\, z_6^2 + 4\, z_1 z_6^2\\
    &+& 
    4\, z_3 z_6^2 - 2\, z_6^3 +  z_2 z_7+  z_1 z_2 z_7 +    9\, z_1^2 z_2 z_7 - 3\, z_2^3 z_7 - 8\, z_2 z_3 z_7 - 
    3\, z_1 z_2 z_3 z_7 + 9\, z_2 z_4 z_7 - 21\, z_5 z_7 \\
    &+ &
    6\, z_1 z_5 z_7 + 4\, z_1^2 z_5 z_7 - 9\, z_3 z_5 z_7 -    6\, z_2 z_6 z_7 + 7\, z_1 z_2 z_6 z_7 + 9\, z_5 z_6 z_7 - 
    9\, z_7^2 + 17\, z_1 z_7^2 - 3\, z_1^2 z_7^2\\
    & -& 
    2\, z_1^3 z_7^2 + 9\, z_1 z_3 z_7^2 - 9\, z_4 z_7^2 +    9\, z_6 z_7^2 - 9\, z_1 z_6 z_7^2) \\
a_{4 5}(z)& =& 2(9\,z_2 - 7\,z_1\,z_2 + 13\,z_1^2\,z_2 - 7\,z_2^3 - 12\,z_2\,z_3 - 7\,z_1\,z_2\,z_3 + 21\,z_2\,z_4 - 28\,z_5 - 7\,z_1\,z_5 + 7\,z_1^2\,z_5 \\
     &- &13\,z_3\,z_5 + 9\,z_4\,z_5 - 3\,z_2\,z_6 + 7\,z_1\,z_2\,z_6 - 3\,z_2\,z_3\,z_6 - 19\,z_5\,z_6 - 4\,z_1\,z_5\,z_6 - 5\,z_2\,z_6^2 + 21\,z_1\,z_7 \\
	&-& 24\,z_1^2\,z_7 + 6\,z_1^3\,z_7 + 7\,z_2^2\,z_7 - 5\,z_1\,z_2^2\,z_7 + 14\,z_3\,z_7 - z_1\,z_3\,z_7 -5\,z_3^2\,z_7 - 19\,z_4\,z_7 + 10\,z_1\,z_4\,z_7 \\
	&-& z_2\,z_5\,z_7 + 10\,z_1\,z_6\,z_7 + 4\,z_1^2\,z_6\,
z_7 + 5\,z_6^2\,z_7 - 2\,z_2\,z_7^2 + 4\,z_1\,z_2\,z_7^2 + 9\,z_5\,z_7^2 - 9\,z_1\,z_7^3)\\
a_{4  6} (z) &=& 2(20 + 12\, z_1 - 18\, z_1^2 + 12\, z_1^3 + 7\, z_2^2 - 6\, z_1\, z_2^2 + 12\, z_3 - 24\, z_1\, z_3 - 6\, z_3^2 + 2\, z_4 \\
	&+& 12\, z_1\, z_4 - \, z_2\, z_5 + 2\, z_6+ 8\, z_1\, z_6 + 2\, z_1^2\, z_6 + 8\, z_3\, z_6 + 6\, z_4\, z_6 - 4\, z_6^2 - 6\, z_2\, z_7 - \, z_1\, z_2\, z_7 \\
	&-& 4\, z_2\, z_3\, z_7 - 10\, z_5\, z_7 - 4\, z_1\, z_5\, z_7 - 5\, z_2\, z_6\, z_7- 20\, z_7^2 + 2\, z_1\, z_7^2 + 4\, z_1^2\, z_7^2 - 9\, z_3\, z_7^2 + 9\, z_6\, z_7^2)\\ 
a_{4  7} ( z) &=& 2(-5\, z_2 - 6\, z_1\, z_2 - 5\, z_2\, z_3 - 19\, z_5 - 4\, z_1\, z_5 - 5\, z_2\, z_6 - 19\, z_7 + 11\, z_1\, z_7 + 5\, z_1^2\, z_7 - 10\, z_3\, z_7\\ 
	&+& 3\, z_4\, z_7 + 9\, z_6\, z_7)\\ 
a_{5  5} ( z) &=& -60 + 48\, z_1 - 12\, z_1^2 - 12\, z_1\, z_2^2 - 24\, z_3 + 24\, z_1\, z_3 - 12\, z_3^2 - 48\, z_4 + 24\, z_1\, z_4 + 12\, z_2\, z_5 \\
	&+& 15\, z_5^2 - 52\, z_6 + 24\, z_1^2\, z_6 - 48\, z_3\, z_6 - 4\, z_4\, z_6 - 48\, z_6^2 - 16\, z_1\, z_6^2 + 40\, z_2\, z_7 + 8\, z_1\, z_2\, z_7 \\
	&-& 8\, z_2\, z_3\, z_7+ 20\, z_5\, z_7 + 8\, z_1\, z_5\, z_7 - 24\, z_2\, z_6\, z_7 - 16\, z_7^2 + 4\, z_1\, z_7^2 - 12\, z_1^2\, z_7^2 + 32\, z_3\, z_7^2 + 28\, z_6\, z_7^2\\ 
a_{5  6} ( z) &=& 2(7\, z_2 - \, z_1\, z_2 - 5\, z_2\, z_3 + 8\, z_5 + 5\, z_1\, z_5 - 7\, z_2\, z_6 + 5\, z_5\, z_6 - 28\, z_7 + 19\, z_1\, z_7 - 6\, z_1^2\, z_7 \\
	&+& 11\, z_3\, z_7- 3\, z_4\, z_7 - 17\, z_6\, z_7 - 9\, z_1\, z_6\, z_7 - 6\, z_2\, z_7^2 + 9\, z_7^3)\\ 
a_{5  7} (z) &=& -20 + 12\, z_1 - 12\, z_3 - 8\, z_4 - 48\, z_6 - 20\, z_1\, z_6 - 12\, z_2\, z_7+ 5\, z_5\, z_7 + 20\, z_7^2\\ 
a_{6  6} ( z) &=& 4(-14 + 12\, z_1 - 6\, z_1^2 + 12\, z_3 - 2\, z_4 + 2\, z_6 + 2\, z_6^2 - 7\, z_2\, z_7 - \, z_5\, z_7 - 5\, z_7^2 - 5\, z_1\, z_7^2)\\ 
a_{6  7} ( z) &=& 2(-7\, z_2 - 3\, z_5 - 19\, z_7 - 11\, z_1\, z_7 + 2\, z_6\, z_7)\\ 
a_{7  7} (z) &=& -60 - 24\, z_1 - 4\, z_6 + 3\, z_7^2\\ 
\eeqrn

With the explicit expression of $\Delta^1$ at our disposal, we can now try to use the Schr\"{o}dinger equation as an efficient mean to compute particular characters of $E_7$. Given that all these characters are polynomials in the $z$-variables, the Schr\"{o}dinger equation can be solved by applying a systematic procedure, which is suitable to be implemented in a computer program able to carry out symbolic calculations. We propose two alternative methods to find the Schr\"{o}dinger eigenfunctions:
\begin{enumerate}
\item Given a weight $\nu=\sum_{i=1}^7n_i\l_i\in P^+$, let us denote $z^{\bf n}$ (or $z^\nu$) the monomial $z^{\bf n}=\prod_{i=1}^7z_i^{n_i}$; thus $z_i=z^{\l_i}$. The operator $\Delta^{1}$ acting on $z^{\bf n}$ gives 
\beq
\Delta^{1} z^{\bf n}=\sum_{\beta\in\Lambda} S_{\beta,{\bf n}}\, z^{\bf n-\beta}=\ve_{\bf m}(1)z^{\bf m}+\sum_{0\ne\beta\in\Lambda} S_{\beta,{\bf n}}\, z^{\bf n-\beta}\,,
\label{e64}
\eeq
where  $\Lambda$ only includes integral linear combinations of the simple roots with non-negative coefficients   and, of course, in the exponent of (\ref{e64}) we express $\beta$ in the basis of fundamental weights. The eigenfunctions $\bchi_{\bf m}$ can be written as
\[
\bchi_{\bf m}=\sum_{\l \in Q^+_{\bf m}} C_\l z^{\bf m-\l}=z^{\bf m}+\sum_{0\ne\l \in Q^+_{\bf m}} C_\l z^{\bf m-\l}\,,
\]
where again the $\l$ in $Q^+_{\bf m}$ are integral linear combinations of the simple roots with non-negative coefficients such  that  they do not give rise to negative powers of the $z$'s. By substituting in the Schr\"{o}dinger equation $\Delta^1\bchi_{\bf m}=\ve_{\bf m}(1)\bchi_{\bf m}$ we find the iterative formula
\[
C_\l=\frac{1}{\varepsilon_{\bf m}(1)-\varepsilon_{\bf m-\l}(1)}\sum_{0\ne\beta\in\Lambda}S_{\beta,{\bf m}-(\l-\beta)}\, C_{\l-\beta}.
\]
To use this formula in practice, one should take into account the heights of the $\l's$ involved, because each coefficient $C_\l$ can depend only on some of the $C_\nu$ such that ${\rm ht}(\nu)<{\rm ht}(\l)$.

\item The Clebsch-Gordan series for the product $\prod_{i=1}^7z_i^{m_i}$ reads
\[
z_1^{m_1}z_2^{m_2}z_3^{m_3}z_4^{m_4}z_5^{m_5}z_6^{m_6}z_7^{m_7}=\bchi_{\bf m}+\sum_{\beta\in Q_{\bf m}}D_{\beta} \bchi_{\bf m-\beta}.
\]
Here it is not difficult, in each particular case, to elaborate a list with all the elements in $Q_{\bf m}$ (i.e., the integral dominant weights appearing in the series). Furthermore, the operator $\Delta^{1}-\varepsilon_{\bf n}(1)$ annihilates the character $\bchi_{\bf n}$. Having this into account, we can make use of the simple-looking formula
\[
\bchi_{\bf m}=\Big\{\prod_{\beta \in Q_{\bf m}}\left(\Delta^{1}-\varepsilon_{\bf m-\beta}(1)\right)\Big\} z^{\bf m}
\]
to obtain the eigenfunctions.
\end{enumerate}
Through any of these methods, it is possible to compute the characters rather quickly. As an illustration, we offer a list of the third order characters in the Appendix A.

Once we have a method for the computation of the characters, we can extend it to produce an algorithm for calculating the Clebsch-Gordan series. Suppose that we want to obtain the series for $\bchi_{\bf m}\cdot\bchi_{\bf n}$. We  list the possible dominant weights entering in the series arranged by heights
\[
\bchi_{\bf m}\cdot\bchi_{\bf n}=\bchi_{\bf m+n}+N_{\mu_1} \bchi_{\mu_1}+N_{\mu_2} \bchi_{\mu_2}+\ldots
\]
The multiplicity  $N_{\mu_1}$ is simply the difference between the coefficients of $z^{\bf \mu_1}$ in $\bchi_{\bf m}\cdot\bchi_{\bf n}$ and in $\bchi_{\bf m+n}$. Then, $N_{\mu_2}$ is the difference between the coefficient of $z^{\bf \mu_2}$ in $\bchi_{\bf m}\cdot\bchi_{\bf n}$ and the sum of the corresponding coefficients in $\bchi_{\bf m+n}$ and $\bchi_{\bf \mu_1}$, and so on. As an example, we present in Appendix B a list with all the cubic Clebsch-Gordan series. 

The approach we are describing is also useful to find the general structure of the series for products of some specific types. Let us consider, for instance, series of the type $z_7 \bchi_{n\lambda_7}$ with arbitrary integer $n>0$. The weights of the representation $R_{\lambda_7}$ are given by the linear combinations $\pm(v_i+v_j), i\neq j$ \cite{ov90}. If we expand these weights in the basis of fundamental weights, we see that there are only four whose coefficients for all $\lambda_i$ with $i\neq 7$ are  non-negative: $\lambda_7, \lambda_6-\lambda_7, \lambda_1-\lambda_7$ and $-\lambda_7$. Hence, the form of the series should be
\beq
z_7 \bchi_{0,0,0,0,0,0,n}=\bchi_{0,0,0,0,0,0,n+1}+ a \bchi_{0,0,0,0,0,1,n-1}+b \bchi_{1,0,0,0,0,0,n-1}+c \bchi_{0,0,0,0,0,0,n-1}, \label{ser11}
\eeq
where we have to fix $a$, $b$ and $c$. Now, by solving the Schr\"{o}dinger equation by means of the first of the two methods described above, one finds
\begin{eqnarray*}
\bchi_{0,0,0,0,0,0,n}&=&z_7^{n}+(1-n) z_6 z_7^{n-2}-z_1 z_7^{n-2}+\ldots\\
\bchi_{0,0,0,0,0,1,n-1}&=&z_6 z_7^{n-1}-z_1 z_7^{n-1}+\ldots\\
\end{eqnarray*}
If we substitute this in (\ref{ser11}), we can solve for $a$ and $b$, obtaining $a=b=1$. We can now fix $c$ by adjusting dimensions in (\ref{ser11}). This gives $c=1$.

We list below the series of the form $z_7 \bchi_{n\lambda_k}$ obtained through the same procedure:
\begin{eqnarray*}
z_7 \bchi_{n,0,0,0,0,0,0}&=&\bchi_{n,0,0,0,0,0,1} +\bchi_{n-1,1,0,0,0,0,0} +\bchi_{n-1,0,0,0,0,0,1} \\
z_7 \bchi_{0,n,0,0,0,0,0}&=&\bchi_{0,n,0,0,0,0,1} +\bchi_{0,n-1,1,0,0,0,0} +\bchi_{0,n-1,0,0,0,1,0} +\bchi_{1,n-1,0,0,0,0,0} \\
z_7 \bchi_{0,0,n,0,0,0,0}&=&\bchi_{0,0,n,0,0,0,1} +\bchi_{1,1,n-1,0,0,0,0} +\bchi_{0,0,n-1,0,1,0,0} +\bchi_{1,0,n-1,0,0,0,1} +\bchi_{0,1,n-1,0,0,0,0} \\
z_7 \bchi_{0,0,0,n,0,0,0}&=&\bchi_{0,0,0,n,0,0,1} +\bchi_{0,1,1,n-1,0,0,0} +\bchi_{1,0,0,n-1,1,0,0} +\bchi_{0,1,0,n-1,0,1,0} +\bchi_{0,0,1,n-1,0,0,1}\\ &+&\bchi_{1,1,0,n-1,0,0,0}+\bchi_{0,0,0,n-1,1,0,0} \\
z_7 \bchi_{0,0,0,0,n,0,0}&=&\bchi_{0,0,0,0,n,0,1} +\bchi_{0,0,0,1,n-1,0,0} +\bchi_{1,0,0,0,n-1,1,0} +\bchi_{0,1,0,0,n-1,0,1} +\bchi_{0,0,1,0,n-1,0,0} \\&+&\bchi_{0,0,0,0,n-1,1,0} \\
z_7 \bchi_{0,0,0,0,0,n,0}&=&\bchi_{0,0,0,0,0,n,1} +\bchi_{0,0,0,0,1,n-1,0} +\bchi_{1,0,0,0,0,n-1,1} +\bchi_{0,1,0,0,0,n-1,0} +\bchi_{0,0,0,0,0,n-1,1} \\
\end{eqnarray*}
\section{Conclusions}
In this paper we have shown how the Calogero-Sutherland Hamiltonian for the Lie algebra $E_7$ can be used to compute both Clebsch-Gordan series and characters of that algebra. The treatment we have presented can be applied to the cases of other simple algebras. It can be also extended to deal with the system of orthogonal polynomials based on $E_7$ for general values of the parameter $\kappa$. The way in which this should be done is the subject of a research now in progress and will be published elsewhere.
\section*{Acknowledgements} 
This work has been partially supported by the Spanish 
Ministerio de Educaci\'{o}n y Ciencia under grants BFM2003-02532 (J.F.N.), BFM2003-00936 / FISI (W.G.F. and A.M.P.), and the sabbatical SAB-2003-0256 (A.M.P.).
\section*{Appendix A: List of the characters of $E_7$ of third order.}

\section*{Appendix B: List of the cubic Clebsch-Gordan series.}
\footnotesize
%

%\begin{thebibliography}{AA99}

%\bibitem{ca71}Calogero F., J. Math. Phys. {\bf 12}, 419--436, 1971.

%\bibitem{su72}Sutherland B., Phys. Rev. {\bf A4}, 2019--2021, 1972.

%\bibitem{appl}D'Hoker E. and Phong D.H., hep-th/9912271; Gibbons G.W. and Townsend P.K., Phys. %Lett. {\bf B454}, 187--192, 1999.

%\bibitem{op83}Olshanetsky M.A. and Perelomov A.M.,  Phys. Rep. {\bf 94}, 313--404, 1983.

%\bibitem{la}Lam\'{e} G., Jour. des Math. Pures et Appliqu\'ees.{\bf 2}, 147--183, 1837; {\it %ibid}  {\bf 4}, 126-163, 1839.

%\bibitem{he80}Hermite Ch., Jour. f\"{u}r die Reine Angew. Math. {\bf 89}, 9--18, 1880. 

%\bibitem{di93}Dittrich J., Inozemtsev V.I., J. Phys. {\bf A26}, L753--L756, 1993.

%\bibitem{in95}Inozemtsev V.I., J. Math. Phys. {\bf 37}, 147--159, 1995.

%\bibitem{pe98a}Perelomov A.M., J. Phys. {\bf A31}, L31--L37, 1998.

%\bibitem{prz98}Perelomov A.M., Ragoucy E. and Zaugg Ph., J. Phys. {\bf A31}, L559--L565, 1998.

%\bibitem{pe99}Perelomov A.M., J. Phys. {\bf A32}, 8563--8576, 1999.

%\bibitem{flp01}Garc\'{\i}a Fuertes W., Lorente M., Perelomov A.M., J. Phys. {\bf A34}, 10963--%10973, 2001, math-ph/0110038.

%\bibitem{fp02}Garc\'{\i}a Fuertes W., Perelomov A.M., Theor. Math. Phys., {\bf 131},
%609-611, 2002; math-ph/0201026.

%\bibitem{la89}Lassalle M., C.R. Acad. Sci., ser. I, {\bf 309}, 941--944, 1989.

%\bibitem{st89}Stanley R.P., Adv. Math. {\bf 77}, 76--115, 1989.

%\bibitem{ma95}Macdonald I.G., {\em Symmetric Functions and Hall  Polynomials}, Oxford Univ. %Press, 1995.

%\bibitem{op89}Opdam E., Inv. Math. {\bf 98}, 1--18, 1989.

%\end{thebibliography}

\end{document}